\documentclass[aip,jcp,preprint,graphicx]{revtex4-1}

\draft 

\usepackage{mathrsfs}
\usepackage{amsmath}

\newcommand{\etal}{{\it~et~al.~}}

\usepackage{graphicx}
\usepackage{amsmath}
\usepackage{hyperref}
\usepackage{epstopdf}
\usepackage{url}

\begin{document}


\title{Incoherent population mixing contributions to phase-modulation two-dimensional coherent excitation spectra} 



\author{Pascal~Gr\'egoire}
\affiliation{D\'epartement de physique, Universit\'e de Montr\'eal, C.P.\ 6128, Succursale Centre-ville, Montr\'eal, Qu\'ebec H3C~3J7, Canada}
\author{Ajay~Ram~Srimath~Kandada}
\affiliation{D\'epartement de physique, Universit\'e de Montr\'eal, C.P.\ 6128, Succursale Centre-ville, Montr\'eal, Qu\'ebec H3C~3J7, Canada}
\affiliation{Center for Nano Science and Technology @Polimi, Istituto Italiano di Tecnologia, via Giovanni Pascoli 70/3, 20133 Milano, Italy}
\affiliation{School of Chemistry and Biochemistry, School of Physics, Georgia Institute of Technology, 901 Atlantic Drive, Atlanta, Georgia 30332, United~States}
\author{Eleonora~Vella} 
\affiliation{D\'epartement de physique, Universit\'e de Montr\'eal, C.P.\ 6128, Succursale Centre-ville, Montr\'eal, Qu\'ebec H3C~3J7, Canada}
\author{Richard~Leonelli}
\affiliation{D\'epartement de physique, Universit\'e de Montr\'eal, C.P.\ 6128, Succursale Centre-ville, Montr\'eal, Qu\'ebec H3C~3J7, Canada}
\author{Carlos~Silva}
\email[e-mail: ]{carlos.silva@gatech.edu}
\affiliation{D\'epartement de physique, Universit\'e de Montr\'eal, C.P.\ 6128, Succursale Centre-ville, Montr\'eal, Qu\'ebec H3C~3J7, Canada} 
\affiliation{School of Chemistry and Biochemistry, School of Physics, Georgia Institute of Technology, 901 Atlantic Drive, Atlanta, Georgia 30332, United~States}


\date{\today}

\begin{abstract}
We present theoretical and experimental results showing the effects of incoherent population mixing on two-dimensional (2D) coherent excitation spectra that are measured via a time-integrated population and phase-sensitive detection. The technique uses four collinear ultrashort pulses and phase modulation to acquire two-dimensional spectra by isolating specific nonlinear contributions to the photoluminescence or photocurrent excitation signal. We demonstrate that an incoherent contribution to the measured lineshape,  
arising from nonlinear population dynamics over the entire photoexcitation lifetime, 
generates a similar lineshape to the expected 2D coherent spectra 
in condensed-phase systems. In those systems, photoexcitations are mobile such that inter-particle interactions are important on any timescale, including those long compared to the 2D coherent experiment. Measurements on a semicrystalline polymeric semiconductor film at low temperature show that, in some conditions in which multi-exciton interactions are suppressed, the technique predominantly detects coherent signal and can be used, in our example, to extract homogeneous linewidths. The same method used on a lead-halide perovskite photovoltaic cell shows that incoherent population mixing of mobile photocarriers can dominate the measured signal since  carrier-carrier bimolecular scattering is active even at low excitation densities, which hides the coherent contribution to the spectral lineshape. In this example, the intensity dependence of the signal matches the theoretical predictions over more than two orders of magnitude, confirming the incoherent nature of the signal. While these effects are typically not significant in dilute solution environments, we demonstrate the necessity to characterize, in condensed-phase materials systems,  the extent of nonlinear population dynamics of photoexcitations (excitons, charge carriers, etc.) in the execution of this powerful population-detected coherent spectroscopy technique. 
\end{abstract}

\pacs{}

\maketitle 
\section{Introduction}

Multidimensional coherent optical spectroscopy has emerged as a powerful tool to unravel various electronic and vibrational coherent phenomena, to separate homogeneous and inhomogeneous linewidth contributions, and to dissect excitation dynamics with intricate detail beyond that routinely afforded by unidimensional ultrafast spectroscopic methods.~\cite{Cho2008,Branczyk2014,Fuller2015,Nardin2016} 
Numerous experimental implementations of 2D coherent spectroscopy have been developed over the last decade, each with unique advantages and disadvantages, providing the ultrafast spectroscopy community an extensive menu of experimental schemes.~\cite{Fuller2015} In this work, we focus on population-detected 2D Fourier transform spectroscopy via phase modulation. This technique, which was pioneered by the group of Andy Marcus a decade ago,~\cite{Tekavec2007} uses four collinear phase-modulated femtosecond laser pulses and implements phase-synchronous detection of a fourth-order population via a time-integrated excitation observable -- the photoluminescence (PL),~\cite{Tekavec2007,Perdomo-Ortiz2012,Widom2013,Bruder2015a,Gregoire2017} photocurrent (PC),~\cite{Nardin2013,karki2014,Vella2016} or photoinduced absorption (PIA)~\cite{Li2016b} signal. 
Because this technique uses lock-in detection using well-defined phase reference waveforms, it provides the formidable advantages that it is insensitive to mechanical and phase instabilities, and that 1/$f$ noise is intrinsically suppressed, allowing extraordinarily high signal-to-noise ratios. Furthermore, unlike four-wave-mixing implementations of 2D coherent spectroscopy, the phase of the response is uniquely defined by the lock-in amplifier, which eliminates the tricky requirement to properly phase the raw data. Here, we address one important potential drawback to this technique when implemented to probe condensed-matter systems in which the measured nonlinear population is subject to nonlinear recombination dynamics, even if these dynamics occur over much longer timescales than those probed by the 2D coherent spectral measurement. Nonlinear \emph{incoherent} population dynamics can produce a signal with 2D lineshape that is \textit{a priori} difficult to distinguish from the desired coherent excitation lineshape. While these processes are typically not present when studying dilute solutions of chromophores as is often the case in many chemical physics and biophysics implementations, these may be dominant in condensed-phase systems such as semiconductors, where the photoexcitations producing the time-integrated response can be mobile and susceptible to multi-particle interactions such as exciton bimolecular annihilation, Auger recombination, and other photocarrier scattering processes. 

Recent theoretical and experimental results, applying the phase-synchronous detection method to coherent multi-photon absorption, have highlighted the role of the excitation signal relaxation mechanisms on the detected nonlinear signal.~\cite{Karki2016,Osipov2016a,Osipov2017} Osipov\etal have shown that cumulative effects, due to remaining excited population from previous laser pulses, strongly affect the amplitude of the measured nonlinear signal when the relaxation time is comparable to the laser repetition rate.~\cite{Osipov2016a,Osipov2017} In this work, we have used a similar theoretical approach to identify an important incoherent contribution to the two-dimensional (2D) spectra measured with the phase-synchronous method. The four collinear pulses are phase modulated with acousto-optic modulators at chosen radio frequencies $\Omega_i~(i=1,2,3,4)$, effectively resulting in two pulse pairs which are intensity modulated at frequencies $\Omega_{21}=\Omega_{2}-\Omega_{1}$ and $\Omega_{43}=\Omega_{4}-\Omega_{3}$.~\cite{Tekavec2007} Since those intensity modulations are purely sinusoidal and, thus, do not contain any higher harmonic components, a nonlinear coherent signal can be isolated at sideband frequencies $\Omega_{43}\pm\Omega_{21}$. This coherent signal, which originates from the successive coherent interactions of the four pulses with the sample, shows the correlations between the different energy levels and is directly used to construct a 2D spectrum as developed by Tekavec\etal~\cite{Tekavec2007} However, we identify an additional contribution to the sideband frequencies which might arise from incoherent population mixing in the sample. This incoherent contribution comes from the generation of two independent subpopulations by each pair of pulses, which can then interact on, for example, nanosecond timescales by any incoherent inter-particle mechanism such as exciton bimolecular annihilation. This incoherent signal results in a 2D spectral lineshape which is similar to the expected coherent signal and also exhibits a similar quadratic intensity dependence. Notably, this signal does not reflect the nonlinear correlation in the sample and, if not carefully considered, can lead to misinterpretations. Here, we develop the theoretical framework to understand the microscopic origin of this incoherent signal and identify a way to reconstruct it from linear (two pulses) measurements. The 2D spectrum of a P3HT film, a semiconducting polymer, is presented to illustrate conditions where the phase-synchronous detection method succeeds in detecting a nonlinear coherent response. 
We then present work on a lead-halide perovskite photovoltaic cell, where incoherent population mixing completely dominates the sideband frequencies signal, thus making it impossible to extract the true coherent 2D response. 

\section{Method and Experiment}

\begin{figure}
\includegraphics[width=8.5cm]{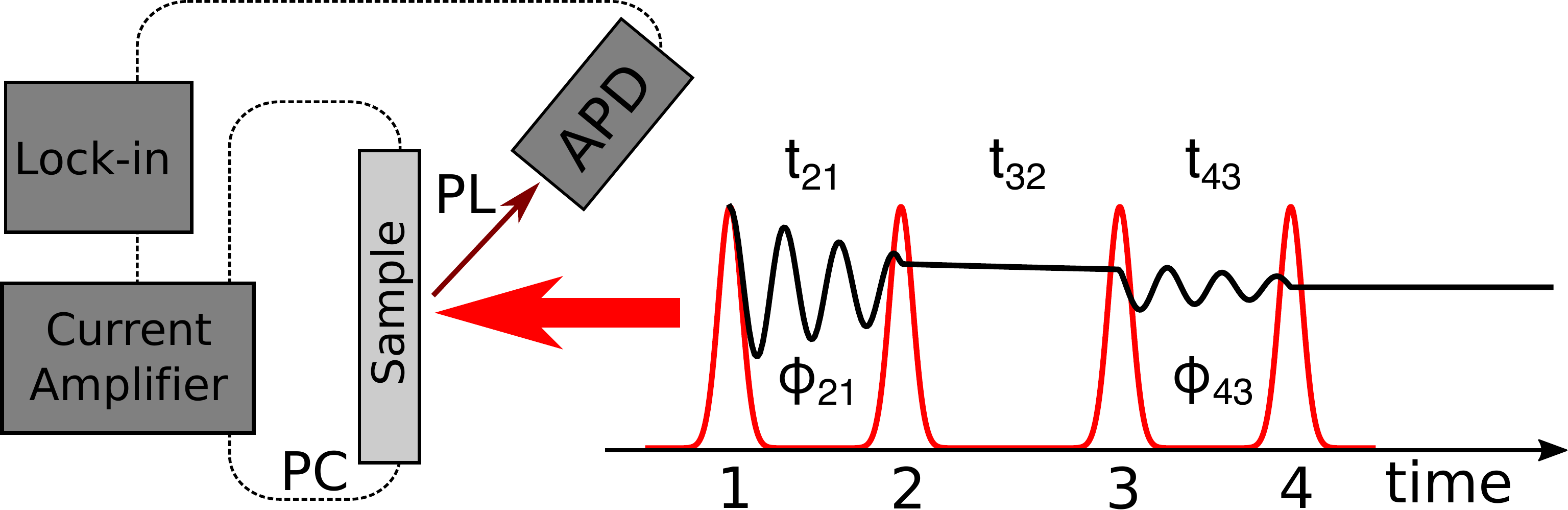}
\caption{Collinear pulse sequence used for 2D coherent spectroscopy measurements. The inter-pulses delays $t_{21}$, $t_{32}$ and $t_{43}$ are controlled by optical delay lines and the relative phase $\phi_{21}$  ($\phi_{43}$) of pulses 1 \& 2  (3 \& 4) is modulated by acousto-optic modulators. For  2D photoluminescence excitation (2D-PLE) spectroscopy, the spectrally-integrated photoluminescence of the sample is collected with an avalanche photodiode (APD), while the 2D photocurrent excitation (2D-PCE) spectroscopy version uses the extracted photocurrent, which is first amplified before being input to the lock-in amplifier (see text for details).\label{fig:setup}}
\end{figure}

The two-dimensional coherent PL excitation (2D-PLE) spectroscopy technique has been developed and described in great detail by Tekavec\etal\cite{Tekavec2007} and Widom\etal\cite{Widom2013} The precise 2D-PLE apparatus used in this work, which is the same for photocurrent excitation (2D-PCE) spectroscopy, have been presented in detail elsewhere.~\cite{Vella2016} In brief, a series of four collinear femtosecond pulses, shown schematically in Fig.~\ref{fig:setup}, are generated using nested Mach-Zehnder interferometers.  The inter-pulse delays $t_{21}$, $t_{32}$ and $t_{43}$ are controlled by three distinct optical delay lines. The phase differences between pulses 1 and 2 ($\phi_{21}$) and between pulses 3 and 4 ($\phi_{43}$) are modulated at frequencies $\Omega_{21}$ and $\Omega_{43}$, respectively, by acousto-optic modulators. The four pulses coherently excite the sample and generate a fourth-order population, which contains the information on the nonlinear susceptibility of the system. This population is detected either with time-integrated PL or, in the case of devices, with photocurrent (see Fig.~\ref{fig:setup}). The phase modulation scheme allows to simultaneously extract the so-called {\it rephasing} and {\it nonrephasing} contributions to the fourth order population by demodulating the signal at sideband frequencies $\Omega_{43}-\Omega_{21}$ and $\Omega_{43}+\Omega_{21}$, respectively, using a lock-in amplifier with optically generated reference waveforms at $\Omega_{43} \mp \Omega_{21}$. To construct a 2D spectrum, the in-phase and in-quadrature signal at both sideband frequencies are measured as a function of the delay times $t_{21}$ and $t_{43}$, corresponding to the {\it coherence} times. The {\it population} time $t_{32}$ is kept constant. Finally, the measured 2D coherence time maps are Fourier Transformed on both $t_{21}$ and $t_{43}$ axes, leading to the rephasing and nonrephasing 2D spectra.

\section{Theory}
\subsection{Linear Absorption Response}

The total electric field is the sum of the four pulses $E(t)=E_1(t)+E_2(t)+E_3(t)+E_4(t)$, with the $j$th pulse ($j=1,2,3,4$) defined as
\begin{equation}
    E_j(t)=A_j(t-t_j)\cos\left[\omega_L(t-t_j)+\phi_j\right],\label{eq:1}
\end{equation}
with $A_j(t-t_j)$ the temporal envelope centred at time $t_j$, $\omega_L$ the laser carrier frequency and $\phi_j$ a constant phase.~\footnote{For the effect of pulse chirp, or spectrally dependent phase, see Tekavec \etal~\cite{Tekavec2006}} The Fourier transform of Eq.~\ref{eq:1} gives the $j$th pulse frequency-dependent field, 
\begin{align}
    &E_j(\omega)=E^{(+)}_j(\omega)+E^{(-)}_j(\omega)\\ 
    &E^{(+)}_j(\omega)=\alpha_j(\omega)e^{i\omega t_j-i\phi_j}~\label{eq:Eplus}\\
    &E^{(-)}_j(\omega)=\alpha_j(-\omega)e^{i\omega t_j+i\phi_j},
\end{align}
where we have defined the forward and counter-rotating terms with $\alpha_j(\omega)$ the spectral amplitude of pulse $j$.



We consider first the effect of only two collinear pulses with an inter-pulse delay $t_{21}$. The starting point for the second order population $n^{(2)}(t)$ created by a pulse pair is the usual perturbative expansion of the density matrix in the interaction picture,~\cite{book:Mukamel,book:Hamm} 
\begin{align}
    n^{(2)}(t)=\left( \frac{i}{\hbar}\right)^2 \int\limits_{0}^{\infty} dt_2 \int\limits_{0}^{\infty} dt_1 E_2(t-t_2) E_1(t-t_2-t_1) R^{(2)}(t_1,t_2).\label{eq:perturbation}
\end{align}
Here we already use the time ordering of pulses 1 and 2. The second-order response function $R^{(2)}(t_1,t_2)$ has explicit form
\begin{equation}
    R^{(2)}(t_1,t_2)=\langle \hat{n}(t_2+t_1) [\hat{\mu}(t_1),[\hat{\mu}(0),\rho]]\rangle,\label{eq:response}
\end{equation}
where the time evolution of operators are  defined as  $\hat{\mathscr{O}}(t)=e^{\frac{i}{\hbar}H_0 t}\hat{\mathscr{O}}e^{-\frac{i}{\hbar}H_0 t}$, $H_0$ being the Hamiltonian of the unperturbed system. In equation~\ref{eq:response}, $\hat{\mu}$ is the dipole operator and $\hat{n}$ is a diagonal operator probing the population in the excited states, which is {\it a priori} proportional to the excitation signal used in the experiment.~\cite{Perdomo-Ortiz2012} Assuming the system is composed of a manifold of excited states $|e\rangle$ coupled to a common ground state $|g\rangle$, $\hat{n}$ has the form
\begin{equation}
\hat{n}(t)=\sum\limits_{e} |e\rangle\langle e|,    
\end{equation}
and the response function reduces to 
\begin{align}
    R^{(2)}(t_1,t_2)
        &= -2 Re \sum\limits_{e} |\mu_{eg}|^2  e^{-i\omega_{eg} t_1}, 
\end{align}
with $\mu_{eg}$ the dipole moment for the transition $g\rightarrow e$ and $\omega_{eg}=\omega_e-\omega_g$ the absorption frequency of the system. 
 Using Eq.~\ref{eq:perturbation} and the rotating wave approximation, the second order population is then
%
\begin{align}
        n^{(2)}(t_{21}) 
    &=\frac{2}{\hbar^2} Re \sum\limits_{e} |\mu_{eg}|^2 \alpha_1(\omega_{eg})\alpha_2(\omega_{eg}) e^{-i \omega_{eg} t_{21}-\Gamma_{eg} t_{21}} e^{i\phi_{21}},\label{eq:n2}
\end{align}
where $\phi_{21}=\phi_{2}-\phi_{1}$ is the phase difference between the two pulses, $t_{21}$ the inter-pulse time delay and $\alpha_j(\omega_{eg})$ the spectral amplitude of the $j$th pulse at frequency $\omega_{eg}$. In Eq.~\ref{eq:n2}, the decoherence of the excited states has been treated phenomenologically by making the substitution $\omega_{eg}\rightarrow \omega_{eg}-i \Gamma_{eg} $. We also supposed that the pulses are short enough such that any pulse overlap effect can be neglected.


\subsection{Phase modulation \label{sec:phmod}}

Acousto-optic modulators enable modulation of the phase of each pulse independently from the inter-pulse delay.~\cite{Tekavec2006} The forward-rotating term of the electric field has then the form
\begin{equation}
    E^{(+)}_{m,j}(\omega)=\alpha_j(\omega)e^{i \omega (t_j+mT)} e^{-i\Omega_j m T},
\end{equation}
where $T$ is the time associated with the repetition rate of the laser, $m$ is an integer counting each laser repetition such that $mT$ is a quasicontinuous time variable and $\Omega_j$ is the modulation frequency of the pulse phase caused by the acousto-optic cell. The $E^{(-)}_{m,j}(\omega)$ term has an analogous form. Comparing with Eq.~\ref{eq:Eplus}, we can assign $\phi_{j}=\Omega_{j} m T$, while the extra phase term $e^{i \omega mT}$, representing the delay between each repetition, cancels out. 

This phase modulation is used to extract the desired response function from the overall signal. For the linear response, the modulation has the form $e^{i\phi_{21}}=e^{i\Omega_{21}mT}$, with $\Omega_{21}=\Omega_2-\Omega_1$, meaning lock-in demodulation with a reference waveform at frequency $\Omega_{21}$ can be used to isolate $n^{(2)}(t_{21})$.~\footnote{The phase modulation procedure has also the effect of {\it down-shifting} the frequency $\omega_{eg}$ while varying $t_{21}$ (see Tekavec\etal~\cite{Tekavec2007}).} The in-phase and in-quadrature components are
\begin{align}
      X^{(2)}(t_{21})&\propto \frac{2}{\hbar^2} Re \sum\limits_{e} |\mu_{eg}|^2 \alpha_1(\omega_{eg})\alpha_2(\omega_{eg}) e^{-i \omega_{eg} t_{21}-\Gamma_{eg} t_{21}},\label{eq:X2_t}\\
      Y^{(2)}(t_{21})&\propto \frac{2}{\hbar^2} Im \sum\limits_{e} |\mu_{eg}|^2 \alpha_1(\omega_{eg})\alpha_2(\omega_{eg}) e^{-i \omega_{eg} t_{21}-\Gamma_{eg} t_{21}},\label{eq:Y2_t}
\end{align}
which can be used to construct the complex signal $\tilde{S}^{(2)}(t_{21})=X^{(2)}(t_{21})+iY^{(2)}(t_{21})$. A Fourier transformation on $t_{21}$ axis gives
\begin{align}
    Re\left[\tilde{S}^{(2)}(\omega_{21})\right]&\propto\frac{2}{\hbar^2}\sum\limits_{e} |\mu_{eg}|^2 \alpha_1(\omega_{eg})\alpha_2(\omega_{eg})
    \frac{\Gamma_{eg}}{(\omega_{21}-\omega_{eg})^2+\Gamma_{eg}^2},\\
    Im\left[\tilde{S}^{(2)}(\omega_{21})\right]&\propto\frac{2}{\hbar^2}\sum\limits_{e} |\mu_{eg}|^2 \alpha_1(\omega_{eg})\alpha_2(\omega_{eg})
    \frac{\omega_{21}-\omega_{eg}}{(\omega_{21}-\omega_{eg})^2+\Gamma_{eg}^2}.
\end{align}
which are proportional to the absorptive and dispersive part, respectively, of the linear susceptibility functions~\cite{Tekavec2006}. Even if the observable $n^{(2)}(t_{21})$ is real-valued, the phase sensitive detection allows to measure the complex-valued response of Eq.~\ref{eq:n2}, which we will denote $\tilde{n}^{(2)}(t_{21})$. 

\subsection{Fourth order coherent population signal~\label{sec:fourthorder}}
A similar development is used by Tekavec\etal to extract the third order nonlinear susceptibility out of four pulses measurements.~\cite{Tekavec2007} The four contributing terms to the fourth order population,  given rise to 2D coherent spectra, have the form~\cite{Tekavec2007}

\begin{equation}
    Z_1 =  \sum\limits_{e,e'} |\mu_{eg}|^2 |\mu_{e'g}|^2 \alpha^2(\omega_{eg})\alpha^2(\omega_{e'g}) e^{-i \omega_{eg} (t_{43}+t_{21})-i(\omega_{eg}-\omega_{e'g})t_{32}} 
    e^{i(\phi_{43}+\phi_{21})},\label{eq:Z1}
\end{equation}
\begin{equation}
     Z_2 =  \sum\limits_{e,e'} |\mu_{eg}|^2 |\mu_{e'g}|^2 \alpha^2(\omega_{eg})\alpha^2(\omega_{e'g}) e^{-i \omega_{eg} t_{43}+i \omega_{e'g} t_{21}-i(\omega_{eg}-\omega_{e'g})t_{32}} 
    e^{i(\phi_{43}-\phi_{21})},\label{eq:Z2}
    \end{equation}

    \begin{equation}
     Z_3 =  \sum\limits_{e,e'} |\mu_{eg}|^2 |\mu_{e'g}|^2 \alpha^2(\omega_{eg})\alpha^2(\omega_{e'g}) e^{-i \omega_{eg} t_{43}+i \omega_{e'g} t_{21} }
    e^{i(\phi_{43}-\phi_{21})},\label{eq:Z3}
    \end{equation}

    \begin{equation}
    Z_4 = \sum\limits_{e,e'} |\mu_{eg}|^2 |\mu_{e'g}|^2 \alpha^2(\omega_{eg})\alpha^2(\omega_{e'g}) e^{-i \omega_{eg} t_{43}-i \omega_{e'g} t_{21}} 
    e^{i(\phi_{43}+\phi_{21})}.\label{eq:Z4}
    \end{equation}
Those terms can be divided into a rephasing and nonrephasing signal

\begin{equation}
n_R^{(4)}(t_{21},t_{32},t_{43})=\frac{2}{\hbar^4}Re(Z_2+Z_3),
\end{equation}
\begin{equation}
n_{NR}^{(4)}(t_{21},t_{32},t_{43})=\frac{2}{\hbar^4}Re(Z_1+Z_4),
\end{equation}
while the total fourth order population is simply the sum of both part
\begin{equation}
    n^{(4)}(t_{21},t_{32},t_{43})=n_{NR}^{(4)}(t_{21},t_{32},t_{43})+n_{R}^{(4)}(t_{21},t_{32},t_{43}).
\end{equation}

\subsection{Intensity dependent excitation signal}

In what precedes, We assumed that the time integrated PL signal analyzed by the lock-in is proportional to the density of the excited state population $n$. However, if intensity dependent mechanisms are present, such as exciton bimolecular annihilation, the nonlinear 2D signal can be affected. As a starting point for our discussion, we describe the excited-state population at times long compared to the femtosecond-laser excitation, by the rate equation
\begin{equation}
    \frac{dn}{dt}=-\gamma n -\beta n^2,\label{eq:rate}
\end{equation}
where the first term is exciton recombination and the second term is exciton-exciton annihilation. The inter-pulse delays and pulse durations are kept much shorter than the population evolution time, typically relaxing on the nanosecond timescale in the semiconductors investigated here.~\cite{DInnocenzo2014a, Paquin2011a} This allows us to consider an instantaneous generation of the initial population density $n_0$, which is created by the pulse sequence, and which oscillates with the imposed phase modulation frequencies $\Omega_{43} \mp \Omega_{21}$. The repetition rate of the laser at 600\,kHz gives a time  $T\sim1.6\,\mu s$ between each pulse sequence, which is much longer than the relaxation time. The effect of previous pulses studied by Osipov\etal~\cite{Osipov2016a,Osipov2017} is then negligible. This situation can be summarized by $t_{21},t_{32},t_{43}\ll 1/\gamma \ll T$.
Using Eq.~\ref{eq:rate}, the time integrated PL intensity $S(n_0)$ has the form
\begin{equation}
    S(n_0)=\frac{\Phi}{\beta}\log \left( 1+\frac{\beta n_0}{\gamma} \right),
\end{equation}
with $\Phi$ accounting for the quantum efficiency of the radiative recombination. Assuming that we work at densities where exciton bimolecular annihilation is weak compared to unimolecular recombination ($\beta n_0 \ll \gamma $),   the closed form expression is
\begin{equation}
    S(n_0)=\frac{\Phi}{\beta}\left[ \frac{\beta n_0}{\gamma} - \frac{1}{2}\left(\frac{\beta n_0}{\gamma}\right)^2 + \dots \right] . \label{eq:S_tay}
\end{equation}
The quadratic term is responsible for the generation of higher harmonics in the measured signal. Indeed, in the case of a two-pulse sequence, the initial density has the form $n_0=n_1+n_2+n^{(2)}(t_{21})$, with $n_1$ and $n_2$ being the unmodulated populations created by only pulse 1 or 2 and $n^{(2)}(t_{21})$ defined as in Eq.~\ref{eq:n2} (neglecting weaker higher order terms). The phase modulation scheme allows to extract the linear signal $S^{(2)}(t_{21})=\frac{\Phi}{\gamma}n^{(2)}(t_{21})$ at modulation frequency $\Omega_{21}$, as expected. However the quadratic term in Eq.~\ref{eq:S_tay} leads also to a term in $(n^{(2)}(t_{21}))^2$ with modulation frequency $2\Omega_{21}$. This second harmonic incoherent signal might have a significant impact in multiphoton experiments, where the modulation technique is used to isolate a coherent signal at higher harmonics of the modulation frequency.~\cite{Bruder2015a,Karki2016,Osipov2016a} The correct nature of those signals is still unclear, whether it is a true multiphoton signal or only the result of non-interacting many-particles~\cite{Mukamel2016}, the incoherent mixing phenomenon can add to the debate, at least when experimenting on condensed-matter systems. 


It is important to point out that the quadratic term, or even higher order terms in $n_0$, are not exclusive to the specific microscopic mechanism chosen here. Many intensity dependent mechanisms, such as Auger processes, photophysical saturation, trapping or amplified spontaneous emission, would conduct to a similar expansion in powers of $n_0$. It is also true if photocurrent is used as an excitation signal, where $\Phi$ can then be understood as the external quantum efficiency of the device.

\subsection{Incoherent mixing in 2D spectra}

For 2D spectroscopy, the excitation is composed of two collinear pairs of pulses with modulation frequencies $\Omega_{21}$ and $\Omega_{43}$. The 2D rephasing and nonrephasing spectra can be measured by demodulating at the sideband frequencies $\Omega_{43}-\Omega_{21}$ and $\Omega_{43}+\Omega_{21}$, respectively.~\cite{Tekavec2007} In this case, up to the fourth order, the density is 
\begin{equation}
    n_0=n^{(2)}(t_{21})+n^{(2)}(t_{43})+n^{(4)}(t_{21},t_{43})+n_{other},
\end{equation}
the first and second term being the linear response from the first and second pulse pair, $n^{(4)}(t_{21},t_{43})$ being the fourth order population generating the coherent rephasing and nonrephasing responses and $n_{other}$ representing all the other terms, either modulated or not, caused by any other interaction combination. Going back to Eq.~\ref{eq:S_tay}, two distinct terms will contribute to the measured signal at the sideband frequencies. First, $S^{(4)}(t_{21},t_{32},t_{43})=\frac{\Phi}{\gamma}n^{(4)}(t_{21},t_{32},t_{43})$ is the coherent signal (see section~\ref{sec:fourthorder}), while an {\it incoherent} signal
\begin{align}
    S^{(inc)}(t_{21},t_{43}) \equiv -\frac{\Phi\beta}{2\gamma^2} \left(n^{(2)}(t_{21}) \times n^{(2)}(t_{43})\right)\label{eq:Inc_short}
\end{align}
comes from the mixing of the linear signals by the $n_0^2$ term. Using Eq.~\ref{eq:n2}, the explicit form is
%
%
\begin{align}
    S^{(inc)}(t_{21},t_{43})=&  -\left(\frac{\Phi\beta}{2\gamma^2} \right)
    \frac{2}{\hbar^4} Re \sum\limits_{e,e'} |\mu_{eg}|^2 |\mu_{e'g}|^2 \alpha(\omega_{eg})^2\alpha(\omega_{e'g})^2 \label{eq:Inc}\\
    &\times (e^{-i (\omega_{eg} t_{21}+ \omega_{e'g} t_{43})} e^{i(\phi_{21}+\phi_{43})}+e^{i (\omega_{eg} t_{21}- \omega_{e'g} t_{43})} e^{i(-\phi_{21}+\phi_{43})}).\nonumber
\end{align}
This expression has a lot of similarities with the $S^{(4)}(t_{21},t_{32},t_{43})$ terms. 
The $e^{i(\phi_{21}+\phi_{43})}$ phase dependent term can be associated with a nonrephasing signal, detected by the lock-in at modulation frequency $\Omega_{43}+\Omega_{21}$, and the $e^{i(-\phi_{43}+\phi_{21})}$ part is similar to a rephasing signal appearing at $\Omega_{43}-\Omega_{21}$. Naturally, $S^{(inc)}$ does not show population beating in $t_{32}$, but the dependencies in $t_{21}$ and $t_{43}$ are the same as the coherent signal, leading to similar peak shapes in the 2D spectra after the Fourier transformation. 

\subsection{Incoherent signal reconstruction from linear signals}

Without quantitative knowledge about the physical mechanisms leading to incoherent mixing (here, values of $\Phi$, $\beta$ and $\gamma$), the absolute amplitude of $S^{(inc)}$ can not be easily evaluated. It is however possible to reconstruct the exact shape of $S^{(inc)}$ using Eq.~\ref{eq:Inc_short}, since both linear signals $n^{(2)}(t_{21})$ and $n^{(2)}(t_{43})$ can be measured during the acquisition of a 2D spectrum using two extra lock-in demodulators referenced at frequencies $\Omega_{21}$ and $\Omega_{43}$. A 2D spectrum is constructed by sampling $t_{21}$ and $t_{43}$ interpulses delays, so no additional manipulations are required to extract those linear signals. 

 Considering the phase detection method (Section~\ref{sec:phmod}), we define reconstructed (normalized) incoherent signals
 \begin{align}
    \tilde{S}^{(inc)}_{NR}(t_{21},t_{43}) \propto \tilde{S}^{(2)}(t_{21}) \times \tilde{S}^{(2)}(t_{43}) \label{eq:Sinc_NR}\\
    \tilde{S}^{(inc)}_{R}(t_{21},t_{43}) \propto \left[\tilde{S}^{(2)}(t_{21})\right]^* \times \tilde{S}^{(2)}(t_{43})\label{eq:Sinc_R},
\end{align}
where $R$ and $NR$ are, in analogy with the coherent signal, the terms detected at modulation frequencies $\Omega_{43}-\Omega_{21}$ and $\Omega_{43}+\Omega_{21}$, respectively. As seen from Eq.~\ref{eq:Inc}, those two definitions are taking into account the different signs in the phase evolution during delays $t_{21}$ and $t_{43}$, which are going to affect the respective shape of $\tilde{S}^{(inc)}_{R}$ and $\tilde{S}^{(inc)}_{NR}$ in the frequency domain.

\subsection{Three-level system}

\begin{figure}
\includegraphics[width=8.5cm]{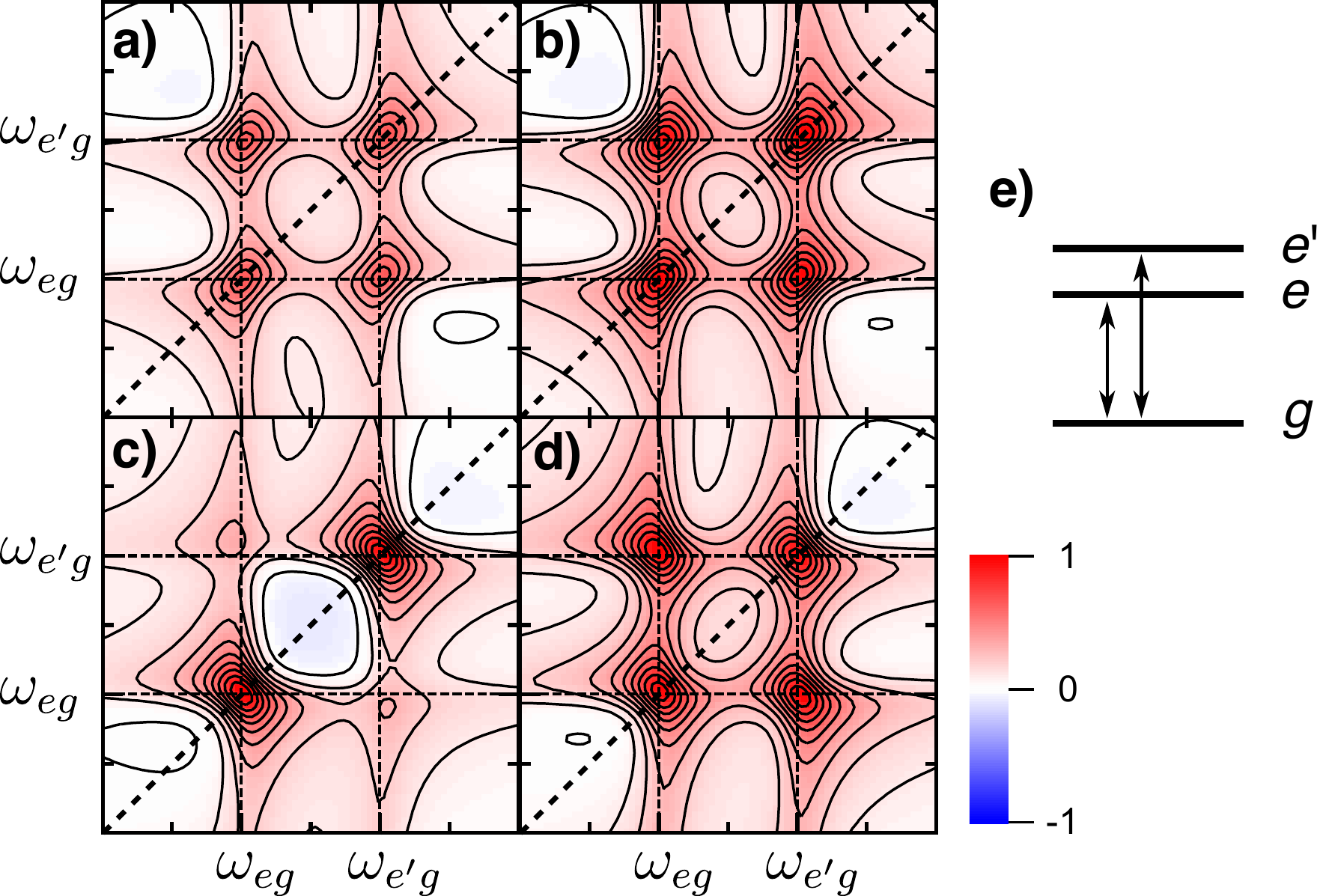}
\caption{Simulated real part of (a) rephasing coherent, (b) rephasing incoherent, (c) nonrephasing coherent and (d) nonrephasing incoherent spectra for the 3-level system shown in (e), with homogeneous broadening $\Gamma=0.2\times\omega_{e'e}$. For the coherent spectra, the population time is set to $t_{32}=0$.\label{fig:3levels}}
\end{figure}

Assuming an ideal homogeneous three-level system, as shown in Fig.~\hyperref[fig:3levels]{\ref*{fig:3levels}(e)}, we can compute the expected 2D spectra for the coherent signals (Equations \ref{eq:Z1}-\ref{eq:Z4}) and the incoherent signals (Eq.~\ref{eq:Inc}). Figures~\hyperref[fig:3levels]{\ref*{fig:3levels}(a)-(b)} compare those two cases for a rephasing signal, while Fig.~\hyperref[fig:3levels]{\ref*{fig:3levels}(c)-(d)} show the nonrephasing part. The latter has clear differences in the off-diagonal peaks, but the former are remarkably similar. For the coherent spectra, relaxation and population beating between the two excited states modify the spectral shape when varying the population waiting time $t_{32}$, while the incoherent signal remains constant.

\subsection{Inhomogeneous broadening}

Fundamental differences between the coherent and incoherent spectra appear when inhomogeneous broadening is considered. Indeed, the power of 2D coherent spectroscopy is to reveal the correlations between energy levels and to overcome inhomogeneously broadened ensembles to expose the homogeneous broadening. The incoherent part does not contain any information about coherent correlations, it is only a result of linear absorption (a product of linear terms as in Equations~\ref{eq:Sinc_NR}-\ref{eq:Sinc_R} ), meaning its shape cannot reveal the homogeneous broadening.

\begin{figure}
\includegraphics[width=8.5cm]{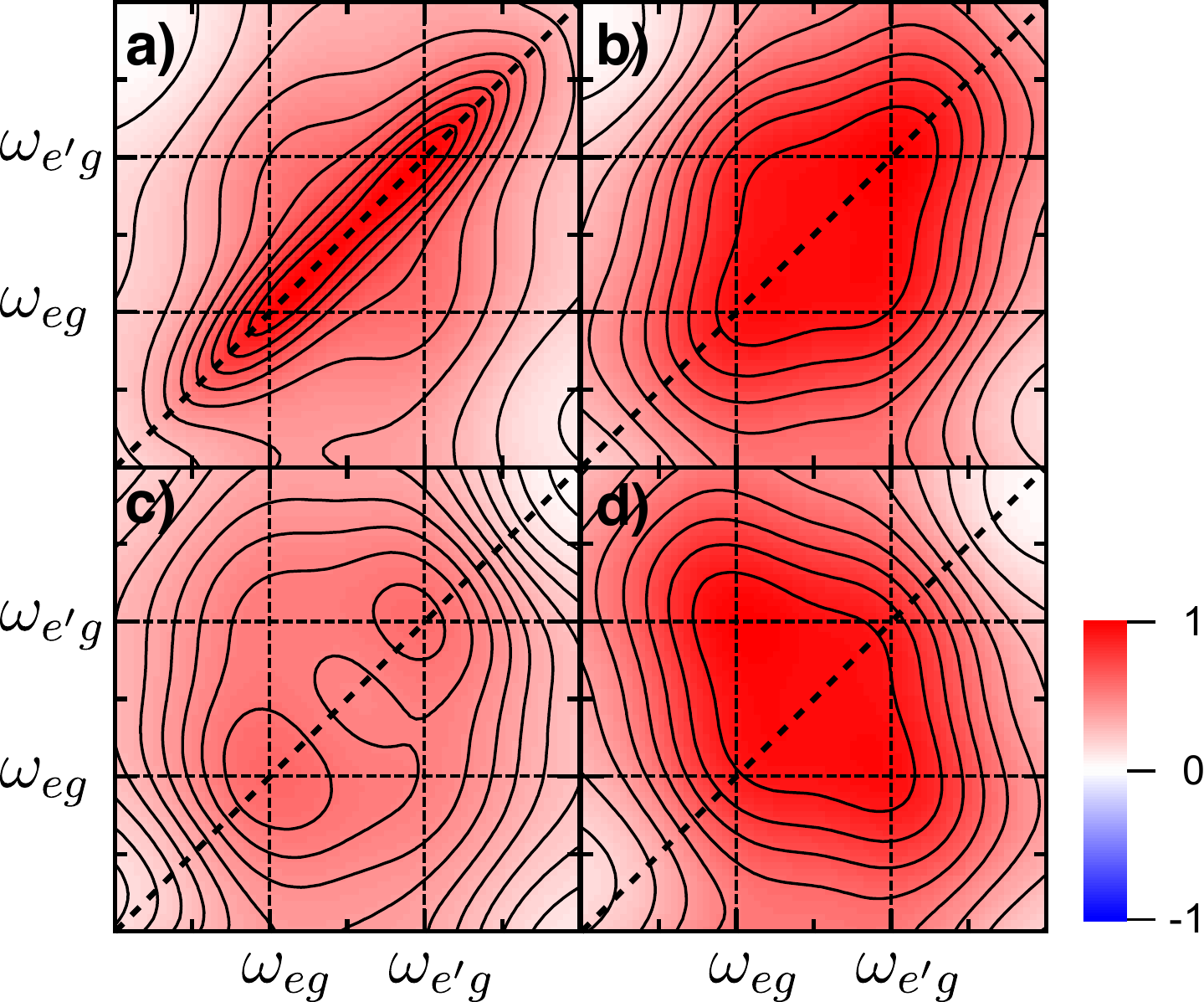}
\caption{Simulation of a inhomogeneously broaden three-level system (see Fig.\hyperref[fig:3levels]{\ref*{fig:3levels}(e)}). The real part of (a) rephasing coherent, (b) rephasing incoherent, (c) nonrephasing coherent and (d) nonrephasing incoherent spectra are shown, using an homogeneous broadening $\Gamma=0.2\times\omega_{e'e}$ and an inhomogeneous Gaussian broadening $\sigma=0.4\times\omega_{e'e}$ (full-width-half-maximum). For the coherent spectra, the population time is set to $t_{32}=0$.\label{fig:3levels_inhomo}}
\end{figure}

The inhomogeneous broadening is included phenomenologically by convolving the response function with a Gaussian distribution for the energy levels. Going back to Eq.~\ref{eq:Inc_short}, the incoherent signal is
\begin{align}
       S^{(inc)}(t_{21},t_{43}) = -\frac{\Phi\beta}{2\gamma^2} \int d\omega_{eg} G(\omega_{eg}-\omega_{eg}^{(0)}) n^{(2)}(t_{21}) \times 
       \int d\omega_{e'g} G(\omega_{e'g}-\omega_{e'g}^{(0)}) n^{(2)}(t_{43}),\label{eq:Inc_short_inhomo}
\end{align}
where $G$ is a Gaussian distribution of width $\sigma$ centered at $\omega_{eg}^{(0)}$, the average transition frequency. In Eq.~\ref{eq:Inc_short_inhomo}, the two convolutions are independent, as opposed to the coherent signal which has terms of the form $\int d\omega_{eg} G(\omega_{eg}-\omega_{eg}^{(0)}) \exp[i (\omega_{eg} t_{21}- \omega_{eg} t_{43})]$ 
in its rephasing part, leading to the well-know photon echo phenomenon, since the transition frequencies during $t_{21}$ and $t_{43}$ are correlated. As shown by simulated spectra of the three-level system, that include inhomogeneous broadening (Fig.~\hyperref[fig:3levels_inhomo]{\ref*{fig:3levels_inhomo}(a)-(b)}), this leads to the characteristic elongated diagonal feature in the rephasing coherent spectrum, which now  differs significantly from its incoherent counterpart. Meanwhile, the simulated nonrephasing spectra are qualitatively similar (Fig.~\hyperref[fig:3levels_inhomo]{\ref*{fig:3levels_inhomo}(c)-(d)}).

\section{Results and discussion}
\subsection{P3HT thin film}

\begin{figure}[ht]
\includegraphics[width=8.2cm]{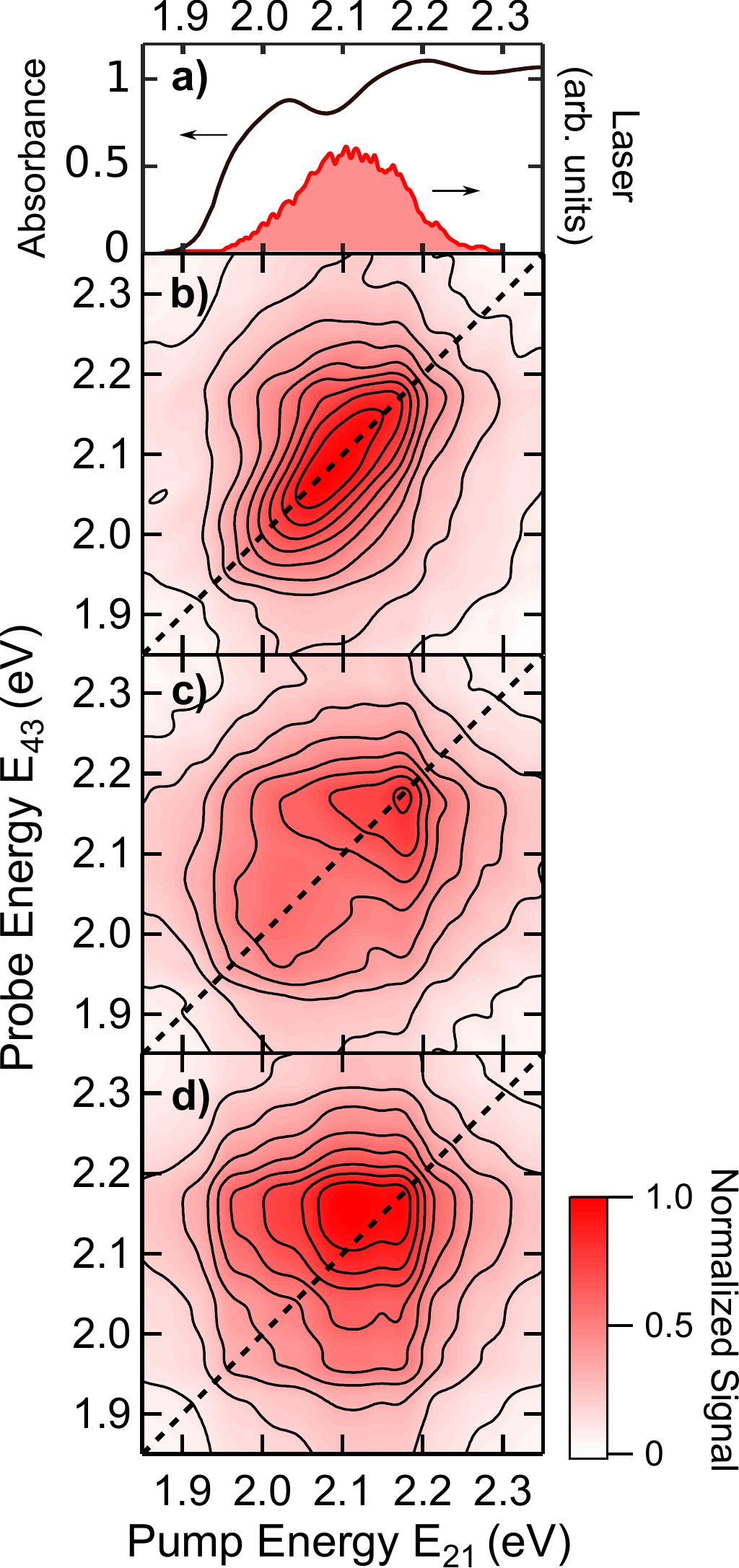}
\caption{(a) P3HT film absorption at 8\,K (black) and spectrum of the femtosecond laser used for excitation (red). Modulus of measured (b) rephasing and (c) nonrephasing 2D spectra on P3HT at 8\,K and with $t_{32}=50$\,fs. (d) Modulus of reconstructed incoherent signal.\label{fig:P3HT1}}
\end{figure}

In a first set of measurements, we employed the 2D-PLE spectroscopy technique on a P3HT thin film at a temperature of 8\,K.~\cite{Gregoire2017} The PL was separated from the laser with a dichroic mirror and sent directly on an avalanche photodiode (APD) (Hamamatsu C12703-01) for spectrally-integrated detection. Fig.~\ref{fig:P3HT1} shows the measured rephasing and nonrephasing 2D spectra detected at modulation frequencies $\Omega_{43}-\Omega_{21}$ and $\Omega_{43}+\Omega_{21}$. In Fig.~\hyperref[fig:P3HT1]{\ref*{fig:P3HT1}(d)}, we also show the incoherent signal $\tilde{S}^{(inc)}_{R}(\omega_{21},\omega_{43})$,~\footnote{While the real and imaginary parts of $\tilde{S}^{(inc)}_{R}(\omega_{21},\omega_{43})$ and $\tilde{S}^{(inc)}_{NR}(\omega_{21},\omega_{43})$ are different, their modulus are the same.} reconstructed from the linear signals at modulation frequencies $\Omega_{43}$ and $\Omega_{21}$ (acquired simultaneously with the nonlinear signals). 

The elongated diagonal feature in the 2D rephasing spectrum is a clear signature of the coherent origin of the measured signal. Indeed, this photon-echo feature cannot be produced by incoherent mixing, where both frequency axes are uncorrelated. As seen in our simulation in Fig.~\ref{fig:3levels_inhomo}, the width of the peak along the diagonal represents the inhomogeneous broadening, here limited by the laser bandwith, and the antidiagonal width shows the homogeneous broadening.~\cite{Tokmakoff2000,Siemens2010} 
The measured nonrephasing spectrum seems also consistent with our simulation (see also ref.~\citenum{Gregoire2017}) and has a clear extra feature on the diagonal at $\sim 2.17$\,eV compared with the reconstructed incoherent spectrum (Fig.~\hyperref[fig:P3HT1]{\ref*{fig:P3HT1}(d)}).

It is useful to examine  the conditions in which the measured 2D-PLE measurement does not reflect significant incoherent mixing contributions. Photoexcitations in P3HT at low temperature are not sufficiently mobile for bimolecular annihilation effects to be significant over their lifetime~\cite{Paquin2011a} and the measurement is carried well within a linear excitation fluence regime.~\cite{Gregoire2017} Furthermore, all secondary dynamics, principally geminate-charge recombination (recombination of charge pairs produced by dissociation of the same original exciton) follow linear fluence dependence.~\cite{Paquin2011a} Under these conditions, the incoherent nonlinear effects discussed in this paper are not significant.



\subsection{Lead-halide perovskite solar cell}

\begin{figure*}
\includegraphics[width=16cm]{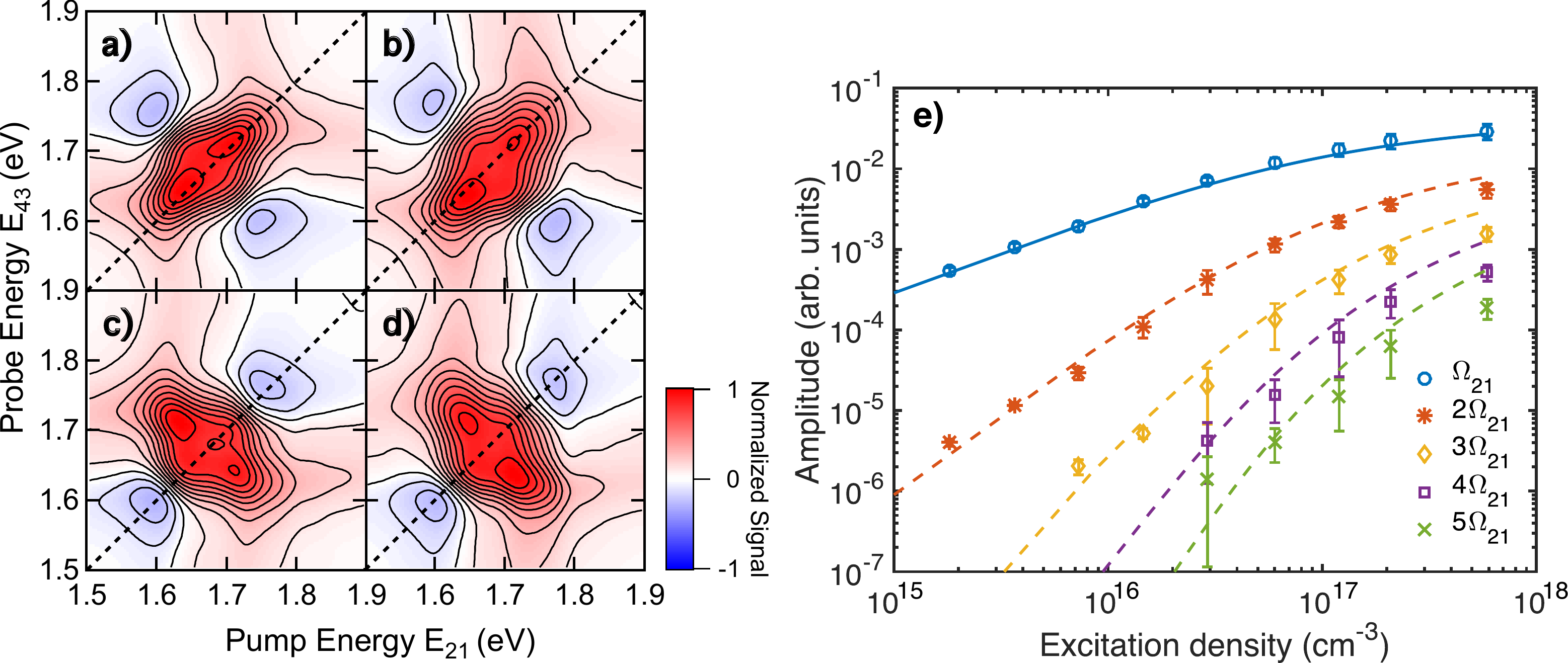}
\caption{Real part of various 2D-PCE spectra on a lead-halide perovskite device at room temperature. (a) Measured 2D rephasing spectrum, (b) reconstructed rephasing incoherent spectrum, (c) measured 2D nonrephasing spectrum and (d) reconstructed nonrephasing incoherent spectrum. (e) Excitation density dependence of perovskite signal at frequencies $m\Omega_{21}$ ($m=1$ to $5$) for a two-pulse experiment. The solid line is a fit of the data with $S(n)=A\,log(1+B\,n)$, where $n$ is the excitation density of carriers, $A=0.023$ and $B=1.32\times10^{-17}$\,cm$^3$. The dashed lines for $m=2$ to $5$ are the calculated incoherent contributions. $n$ is derived from the device thickness and quantum efficiency and from the excitation power density.
 \label{fig:pero1}}
\end{figure*}

In a second set of measurements, we used both photocurrent (PC) and PL detection methods to measure the 2D response of a lead-halide perovskite photovoltaic cell.~\cite{Tao2015} 
The photocurrent was converted to a voltage with a current-to-voltage amplifier (Zurich Instruments HF2TA) before the lock-in demodulation. Fig.~\hyperref[fig:pero1]{\ref*{fig:pero1}(a)-(d)} compare the room temperature 2D-PCE rephasing and nonrephasing signals with the reconstructed incoherent spectra. Apart from small variations attributed to experimental noise, the measured and reconstructed spectra are identical. From this comparison, the sum and difference frequency signal appear to be completely dominated by the incoherent mixing effect. 


\begin{figure}
\includegraphics[width=8.5cm]{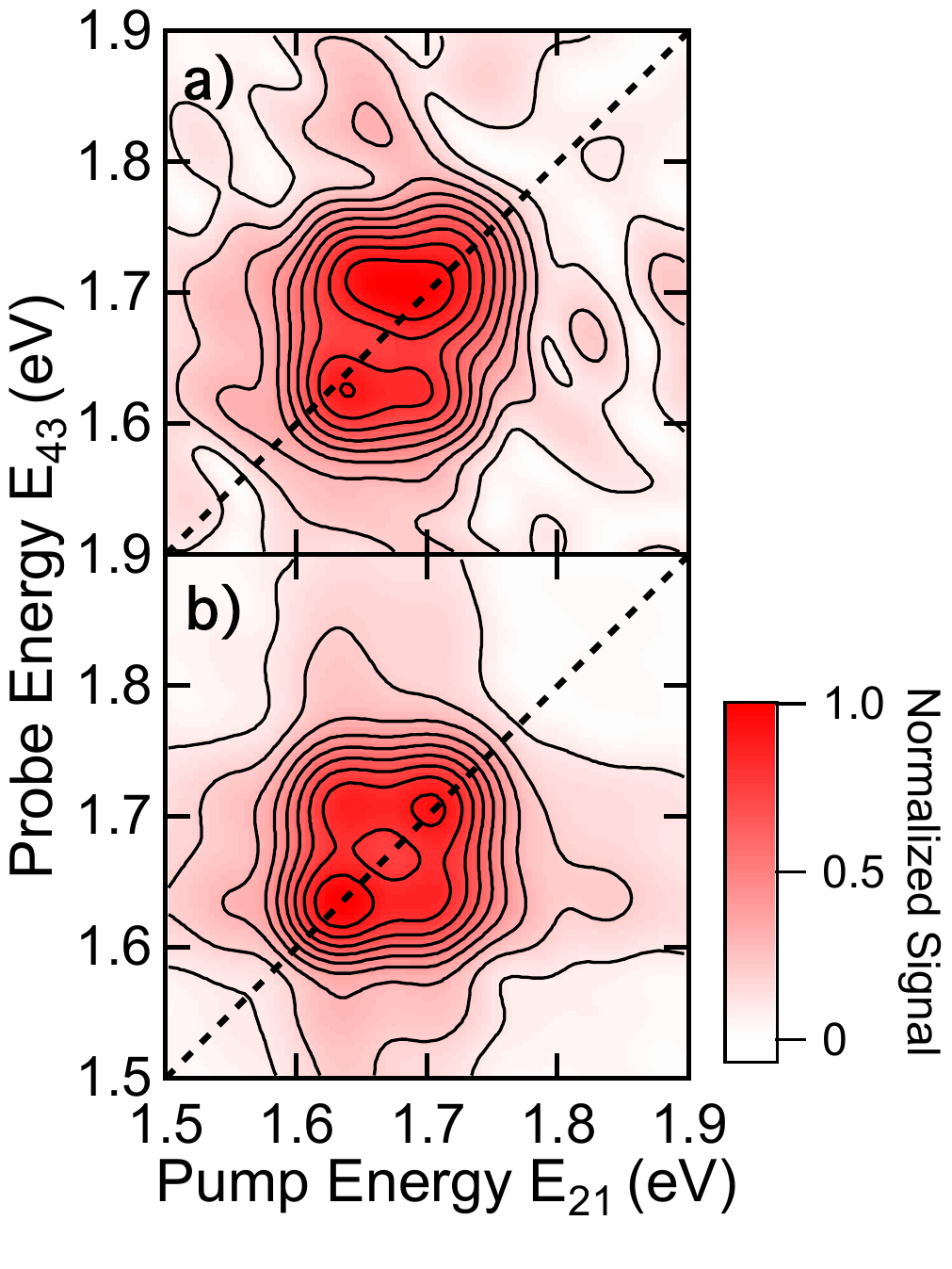}
\caption{ Modulus of nonrephasing (a) 2D-PLE (b) 2D-PCE spectra of a lead-halide perovskite device measured at 12\,K.\label{fig:peroPCPL}}
\end{figure}

Incoherent nonlinearities may also arise from the detection method via electronic impedances. To eliminate such a contribution, we carried out simultaneous PL and PC measurements. In order to increase the PL quantum yield, which is very low in this device at room temperature, we performed the experiments at 12\,K. As shown in Fig~\ref{fig:peroPCPL}, the PL and PC spectra are comparable, their peak shape are  only limited by the excitation spectrum, suggesting the presence of incoherent mixing in both of these measurements and discarding any electronic impedance effects.

In Fig.~\hyperref[fig:pero1]{\ref*{fig:pero1}(e)}, we simplified our experiment to extract the amplitude of the incoherent signal and compared it with our theoretical predictions. To do so, we used only one pair of pulses and compared the total linear signal detected in photocurrent at $\Omega_{21}$ with the signal at $2\Omega_{21}$. Again, this signal has two possible origins. It could be a coherent signal consisting in a doubly excited state,~\cite{Bruder2015a} where the system interacts twice with each pulse to end up in a doubly excited population. The other possible origin is an incoherent mixing of the linear signal $\tilde{S}^{(2)}(t_{21})$ with itself, caused by the quadratic term of Eq.~\ref{eq:S_tay}, which is exactly the same term responsible for the incoherent mixing at $\Omega_{43}\pm\Omega_{21}$ in 2D spectra. 

Unlike excitonic semiconductors as P3HT, PC and PL here are dependent on the integrated electron ($n$) and hole ($p$) densities: $PL = \int B_{rad} n(t)p(t)dt $ and $PC = \int n(t)dt + \int p(t) dt$. It has been demonstrated in perovskites that carrier recombination follows a trap limited behavior at low excitation densities.\cite{SrimathKandada2016} Assuming electron-only trapping, it can be shown that the steady-state densities of electrons and holes scale as $n \propto n_0$ and $p \propto n_0^{0.5}$. This imposes a nonlinear behavior in both PL and PC, albeit with distinct relationships. For instance, in the case of PC at densities much lower than the available trap density, one would observe only hole current, which  scales as $n_0^{0.5}$. As the density is increased closer to the available defect density, a substantial electron current appears, which scales linearly with excitation density. The total PC is then a combination of linear and sub-linear trends.  At higher densities, the limiting channel is radiative bimolecular recombination, which results again in a $n_0^{0.5}$ scaling. Both these regimes are evident on the data shown in figure~\hyperref[fig:pero1]{\ref*{fig:pero1}(e)} where the $\Omega_{21}$ signal amplitude varies linearly at low excitation power and switches to a sublinear trend at higher density. It must be highlighted that PL at low excitation densities scales as $n_0 ^{3/2}$ due to the product of electron and hole densities, unlike in the case of the PC. Nevertheless, the presence of the nonlinear response will manifest itself as an incoherent signal in the phase-modulated experiment, and this makes incoherent mixing contributions to the 2D measurement unavoidable. 

In Fig.~\hyperref[fig:pero1]{\ref*{fig:pero1}(e)}, we also show the measured signal at $m\Omega_{21}$ ($m=3,4,5$) as a function of the excitation density, originating either from a $m$th coherent excited state  or from higher order terms in the expansion of Eq.~\ref{eq:S_tay}. The data at $\Omega_{21}$ (blue circles) have been used to parametrize the photocurrent excitation density ($n$) dependence with the phenomenological function $S(n)=A\,\log(B\,n)$. The blue solid line is the best fit to the data, with $A=0.023$ and $B=1.32\times10^{-17}$\,cm$^3$. Assuming that incoherent mixing dominates, we simulate the amplitude of the signal at $m=2,3,4,5$ using only $S(n)$ as input, resulting in the dashed lines in Fig.~\hyperref[fig:pero1]{\ref*{fig:pero1}(e)}. 

The goodness of the fit on more than two orders of magnitude in excitation density demonstrate that incoherent mixing dominates the signal at harmonics of $\Omega_{21}$. Together with the strong similarities between the measured 2D spectra at $\Omega_{43}\pm\Omega_{21}$ and the reconstructed signals, we can conclude that the coherent signal is masked by a strongly dominant incoherent background in perovskite 2D spectra. 
In Fig.~\hyperref[fig:pero1]{\ref*{fig:pero1}(e)}, there is still a substantial signal at $2\Omega_{21}$ in the lower carrier density regime, where the signal at $\Omega_{21}$ looks completely linear. The sensitivity of the technique enables detection of even weak nonlinearities in PL and PC, which could be misinterpreted as a coherent signal if the full carrier density dependence is unknown. Furthermore, the $2\Omega_{21}$ and $\Omega_{43}\pm\Omega_{21}$ signals are varying as $I^2$ in the lower density regime, which is also the expected intensity dependence for the coherent signal. It is then unlikely that reducing the excitation density will allow the coherent signal to overcome the incoherent part in this perovskite solar cell. While photocurrent measurements can be exploited with this technique to achieve extraordinarily high signal-to-noise ratios,~\cite{Bakulin2016} the contribution of incoherent mixing effects must always be examined closely.

Accumulation effects suggested by Osipov et al~\cite{Osipov2017} can also be pertinent in the case of perovskite solar cells given the long carrier lifetimes in these systems, that can extend beyond the repetition rate of the laser. It must be noted, however, that most of the carrier recombination is complete within the first few hundreds of nanoseconds. The microsecond lifetime observed previously is associated with the carrier recombination at various interfaces within the device;~\cite{Tao2017} the residual carrier population over these timescales will be substantially smaller than on sub-microsecond timescales, especially with efficient carrier extraction and under femtosecond excitation. Even though such accumulation effects cannot be completely disregarded, we consider that the signal from incoherent population mixing will be the major contributor to the measured non-linear signal in the device data discussed in this work.      

It is important to underline that this conclusion does not mean that 2D-PCE measurements are necessarily dominated by incoherent mixing effects. In particular, we note that in the report by Karki\etal, which discussed ultrafast dynamics in semiconductor-nanocrystal solar cells, these effects do not appear to be dominant.~\cite{karki2014} In that report, comparison of 2D-PLE and 2D-PCE measurements displayed distinctive differences that were exploited in the interpretation of the data. Those differences cannot be interpreted by the incoherent mixing effect discussed here. We consider that when possible, simultaneous 2D-PLE and 2D-PCE measurements can be helpful to examine the role of incoherent population effects in condensed-matter material systems. 

Finally, we consider possible ways to overcome the incoherent population mixing effects discussed here. If one is able to time-gate the population observable such that the cumulative second-order population dynamics are not significant during the detection gate, one would expect that the incoherent mixing contribution would be suppressed. The photoinduced absorption variant of 2D phase-modulation coherent excitation spectroscopy discussed by Li\etal~\cite{Li2016b} is particularly conducive to efficient time-gating of the response when using a femtosecond probe pulse derived, for example, from a white-light continuum. Similarly, time-gating the photoluminescence in the 2D-PLE variant would permit suppression of the incoherent mixing contributions from nonlinear population dynamics over long timescales. These modifications to the detection scheme would add a layer of complexity to this experimental implementation of 2D coherent excitation spectroscopy, but they would permit exploiting the power of probing population observables, comparing photoluminescence, photocurrent, and in fact any population probe, to unravel electronic excitation dynamics in a broad class of condensed-matter systems and operating optoelectronic devices.


\section{Conclusion}

We have demonstrated that incoherent population mixing can contribute to the measured signal in 2D  coherent excitation spectroscopy. This incoherent signal can lead to similar spectral lineshapes as the expected coherent signal, since it originates from a convolution of the linear response on the $\omega_{21}$ and $\omega_{43}$ axes. Careful analysis needs to be carried out to discriminate which of the coherent and incoherent contributions dominate the measurements. We have shown that the coherent signal is prevalent in P3HT 2D-PLE spectra, where the characteristic diagonal feature caused by the interplay between homogeneous and inhomogeneous broadening can only be explained via a coherent interaction. In a lead-halide perovskite solar cell, this coherent signature is not evident and the measured spectra matches the reconstructed incoherent signal. Furthermore, the sublinear photocurrent intensity dependence is in agreement with our modeled intensity dependance of the incoherent mixing term. We can then conclude that the photocurrent nonlinear behavior with intensity is too strong in this kind of sample and prevents the measurement of a coherent 2D spectra. Those two extreme cases, namely fully coherent or incoherent effects, show that knowledge about the excitation signal is crucial to obtain valid 2D spectra using this population-detected phase modulation scheme.The technique is ideally suited to study coherent phenomena in non-interacting chromophores, where the coherently excited population on a given site cannot interact with other excitations before emitting PL or engendering a PC. Applying this techniques to solid-states samples, where excitations are instrinsically mobile, requires knowledge of photoexcitation dynamics over their lifetimes at the density regime under investigation. 




\begin{acknowledgments}
CS, EV and PG are indebted to Prof.\ Andy Marcus and Dr.\ Julia Widom for their enthusiastic assistance in the construction of the experimental setup. CS and RL acknowledge funding from the Engineering and Physical Sciences Research Council of Canada (NSERC), the Canada Foundation for Innovation, and The Fonds du Qu\'ebec -- Nature et Technologies. CS further acknowledges funding from the Universit\'e de Montr\'eal University Research Chair. PG acknowledges funding from NSERC via a Doctoral Postgraduate Scholarship. ARSK acknowledges funding from  EU Horizon 2020 via Marie Sklodowska Curie Fellowship (Global) (Project No. 705874). The authors thank Chen Tao and Annamaria Petrozza for the perovskite solar cell and Matthew Dyson, Natalie Stingelin and Paul~N.~Stavrinou for the P3HT films. 
\end{acknowledgments}


%

\end{document}